\documentclass[english]{IEEEtran}
\usepackage[T1]{fontenc}
\usepackage[latin9]{inputenc}
\usepackage{float}
\usepackage{slashed}
\usepackage{amsmath}
\usepackage{graphicx}
\usepackage{amssymb}

\makeatletter

\floatstyle{ruled}
\newfloat{algorithm}{tbp}{loa}
\floatname{algorithm}{Algorithm}

\newtheorem{prop}{Proposition}
\newtheorem{proof}{Poof}

\makeatother

\usepackage{babel}

\begin{document}

\title{MMSE Based Greedy Antenna Selection\\
Scheme for AF MIMO Relay Systems}

\author{Ming Ding, Shi Liu, Hanwen Luo, and Wen Chen, \textit{Member, IEEE}%
\thanks{Copyright (c) 2008 IEEE. Personal use of this material is permitted.
However, permission to use this material for any other purposes must
be obtained from the IEEE by sending a request to pubs-permissions@ieee.org %
}
\thanks{The authors are with the Department of Electronics Engineering, Shanghai
Jiaotong University, China (E-mail: \{dm2007; liushi\_9851; hwluo;
wenchen\}@sjtu.edu.cn). Wen Chen is also with SEU National Key Lab
for mobile communications.

The authors would like to thank Dr. Sayantan Choudhury from Sharp
Lab of America and several anonymous reviewers for their careful review
on earlier versions of the manuscript. This work is supported by NSFC
60972031, SEU SKL for Mobile communications W200907.%
}}

\maketitle
{}
\begin{abstract}
We propose a greedy minimum mean squared error (MMSE)-based antenna
selection algorithm for amplify-and-forward (AF) multiple-input multiple-output
(MIMO) relay systems. Assuming equal-power allocation across the multi-stream
data, we derive a closed form expression for the mean squared error
(MSE) resulted from adding each additional antenna pair. Based on
this result, we iteratively select the antenna-pairs at the relay
nodes to minimize the MSE. Simulation results show that our algorithm
greatly outperforms the existing schemes.\end{abstract}

\section{Introduction}

Multipe-Input multiple-output (MIMO) relay systems
have been recognized to achieve a large diversity gain and a large
spectrum efficiency \cite{[1]MIMO-relay capacity 2005}. Meanwhile,
amplify-and-forward (AF) MIMO relay systems have drawn extensive attentions
in the literature due to their simplicity and mathematical tractability
\cite{[1]MIMO-relay capacity 2005}\textendash{}\cite{[9]S-O}. In
order to achieve the theoretical capacity shown in \cite{[1]MIMO-relay capacity 2005,[2]scaling law 2006},
many advanced signal processing schemes have been proposed. The authors
of \cite{[2]scaling law 2006} introduced the \textquotedblleft{}doubly
coherent\textquotedblright{} backward and forward matched filtering
strategy. In \cite{[3]MF-ZF 2008}, backward matched filtering and
forward zero-forcing (ZF) precoded transmission was proposed. In \cite{[4]SVD-ZF-DPC 2008},
singular value decomposition (SVD) for backward channel and ZF dirty
paper coding at the source node was investigated.

However, most prior work entails that relay nodes should be equipped
more antennas than the source or destination node so that the relays
can perform backward interference-suppressing reception and forward
precoding function. Unfortunately, this is not a realistic assumption
since in practice, relays basically serve as a low-cost and low-complexity
means to extend the coverage and enhance the spectrum efficiency for
cell edge users \cite{[5]cellular relay}. Therefore, in the practical
cases that advanced signal processing cannot be relegated to the relay
nodes, relay/antenna selection becomes an attractive option.

A few antenna selection techniques designed for single antenna/stream
relay networks \cite{[6]single antenna relay network,[7]single stream relay network}
have been reported recently. Many on-going works regarding antenna
selection for multi-stream AF MIMO relay systems unfold more interesting
thoughts. In \cite{[8]DORS}, a heuristic relay/antenna selection
criterion based on harmonic mean of dual-hop sub-channel gains was
proposed. In \cite{[9]S-O}, the authors proposed an iterative antenna
selection scheme based on semi-orthogonality among the selected antenna
pairs. However, both \cite{[8]DORS} and \cite{[9]S-O} overlooked
the noise at the relay nodes, which might be too ideal in practice.
Moreover, instead of using heuristic methods, it would be better to
develop an antenna selection scheme based on more concrete criteria
in closed forms, such as capacity maximization or minimum mean squared
error (MMSE). Thus, in this letter we propose a MMSE-based greedy
antenna selection algorithm for AF MIMO relay systems. Simulation
results validate the superiority of our scheme compared to those in
\cite{[8]DORS} and \cite{[9]S-O}, with the gain being more pronounced
when the noise at the relay nodes is relatively large.

In this letter, $\left(\cdot\right)^{\textrm{T}}$,$\left(\cdot\right)^{\textrm{H}}$,$\left(\cdot\right)^{\textrm{-1}}$,
$\textrm{det}\left(\cdot\right)$, and $tr\left(\cdot\right)$ stand
for the transpose, conjugate transpose, inverse, determinant and trace
of a matrix, respectively. $\varepsilon\left(\cdot\right)$ is the
expectation of a random variable. $\left|\mathbf{a}\right|$ denotes
the Euclidean norm of a vector $\mathbf{a}$. $\mathbf{\left(A\right)}_{i,i}$
is the $\mathit{i}$th diagonal entry of the matrix $\mathbf{A}$.
$\mathbf{I}_{N}$ stands for an $N\times N$ identity matrix. Finally,
$\textrm{C}_{K}^{l}$ counts the events of selecting $l$ elements
from a homogeneous set containing $K$ elements.

\section{System Model }

The dual-hop MIMO relay system considered in this letter is illustrated
by Fig.\,1, where the source node ($\mathcal{S}$), the destination
node ($\mathcal{D}$) and each relay node $\mathcal{R}_{k}$ $\left(k=1,\cdots,K\right)$
are equipped with $N_{\textrm{s}}$, $N_{\textrm{d}}$, and $N_{\textrm{r}}$
antennas, respectively.  ${\mathbf{H}}_{k}\in\mathbb{C}^{\mathit{N_{\textrm{r}}\times N_{\textrm{s}}}}$
and ${\mathbf{G}}_{k}\in\mathbb{C}^{\mathit{N_{\textrm{d}}\times N_{\textrm{r}}}}$
denote the backward channels ($\mathcal{S\rightarrow\mathcal{R_{\mathit{k}}}}$)
and the forward channels ($\mathcal{\mathcal{R_{\mathit{k}}\rightarrow D}}$)
respectively. All the channels are modeled as block-wise flat fading.
Throughout the letter, we only focus on Half Time Division Duplex
(HTDD) relaying with AF protocol, \textit{i.e.}, the transmission
time interval is divided into 2 time slots. The first and the second
time slot of which are assigned to the backward and the forward transmission
respectively.

\begin{figure}[H]
\centering \includegraphics[width=8cm,height=3.8cm]{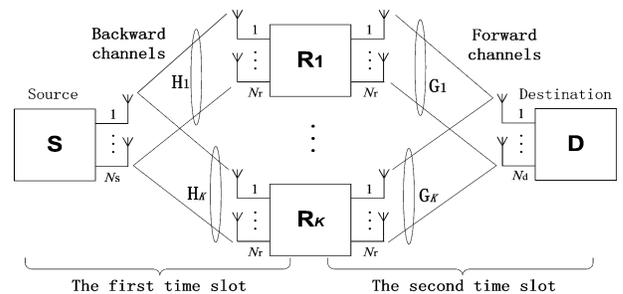}
\vspace{0.5em}
\renewcommand{\figurename}{Fig.}\caption{Schematic model of an AF MIMO relay system.}

\end{figure}

Taking the issues of practical implementation into consideration,
we here assume that each MIMO relay node is equipped with only one
power amplifier (PA) so that only one antenna pair can be activated
on each relay node for the HTDD based transmission. We will select
$\mathit{L}$ antenna pairs in the MIMO relay network to bridge the
communication from $\mathcal{S}$ to $\mathcal{D}$. Let the $\mathit{L}$
antenna pairs be associated with a relay set $\Gamma=\left\{ r\left(l\right)|l=1,\cdots,L\right\} $,
which contains distinct elements due to the single-PA assumption.
It should be pointed out that the single-PA restriction can be easily
lifted to multi-PA by using the eigenmode construction \cite{[9]S-O},
\textit{i.e.}, SVD operations are performed for $\mathbf{H_{\mathit{k}}}$,
$\mathbf{G_{\mathit{k}}}$, resulting in eigenmode-based definitions
for the equivalent backward and forward channels. Here we consider
a case $K\geq N_{\textrm{s}},N_{\textrm{d}}$. Thereby, the multiplexing
gain of the interested system is limited by $M=\textrm{min}\left(N_{\textrm{s}},N_{\textrm{d}}\right)$.
If full multiplexing gain is achieved between $\mathcal{S}$ and $\mathcal{D}$,
$\mathit{L}$ should be no less than $M$ in order to make the equivalent
two-hop relay channel well-conditioned. In \cite{[8]DORS} and \cite{[9]S-O},
$\mathit{L}$ is set to $M$ due to the dimensional condition of the
orthogonal sub-channels, whereas in our scheme $\mathit{L}$ is encouraged
to be larger than $M$ so that more diversity could be exploited to
benefit the MSE performance. For simplicity, we assume $N_{\textrm{d}}\geq N_{\textrm{s}}=M$
in the sequel without loss of generality.

The $\mathit{L}$ antenna pairs generate a compound backward channel
$\mathbf{H=\left[{\mathbf{h}}_{\left(\mathit{b}(\textrm{1}),\mathit{r}(\textrm{1})\right)}^{\textrm{T}},\cdots,{\mathbf{h}}_{\left(\mathit{b}(\mathit{l}),\mathit{r}(\mathit{l})\right)}^{\textrm{T}},\cdots,{\mathbf{h}}_{\left(\mathit{b}(\mathit{L}),\mathit{r}(\mathit{L})\right)}^{\textrm{T}}\right]^{\textrm{T}}}$,
where ${\mathbf{h}}_{\left(\mathit{b}(l),\mathit{r}(l)\right)}^{\textrm{}}$
is the $1\times N_{\textrm{s}}$ channel vector of $\mathcal{S}$
to the $\mathit{l}$th selected backward antenna $b\left(l\right)$
on relay $r\left(l\right)$. During the first time slot, let the received
signals at the $L$ backward antennas be stacked into a vector $\mathbf{y=\left[\mathit{y_{\textrm{1}},\mathit{y_{\textrm{2}},\cdots,\mathit{y_{L}}}}\right]^{\textrm{T}}}$,
which is given by\begin{equation}
\mathbf{y}=\mathbf{Hx}+{\mathbf{n}}_{\textrm{r}}^{\,},\end{equation}

\noindent where $\mathbf{x\in\mathbb{C\mathit{^{N_{\textrm{s}}}}}}$
is the transmit signal vector, and ${\mathbf{n}}_{\textrm{r}}^{\,}\in\mathbb{C\mathit{^{L}}}$
denotes the white zero-mean circularly symmetric complex Gaussian
(ZMCSCG) noise vector with covariance matrix ${\mathbf{I}}_{L}$.
We further denote by $\mathit{P_{\textrm{s}}}$ the total transmit
power available at $\mathcal{S}$ and put constraint on $\mathbf{x}$
as

\begin{equation}
\varepsilon\left\{ tr\left(\mathbf{xx^{\textrm{H}}}\right)\right\} =tr\left(\varepsilon\left\{ \mathbf{xx^{\textrm{H}}}\right\} \right)\leq\mathit{P_{\textrm{s}}}.\end{equation}

\noindent To derive tractable solutions, we assume equal power allocation
across the transmitted data streams at $\mathcal{S}$ with full power.
Hence, the covariance matrix of $\mathbf{x}$ is $\sigma_{x}^{2}{\mathbf{I}}\mathbf{_{\mathit{N_{\textrm{s}}}}}$,
and $\sigma_{x}^{2}=\mathit{P_{\textrm{s}}}/N_{\textrm{s}}$.

In the second time slot, relays in $\Gamma$ amplify and forward the
received signal $\mathbf{y}$ to $\mathcal{D}$. We assume that no
direct link is available from $\mathcal{S}$ to $\mathcal{D}$ due
to long-distance pathloss. The AF relays\textquoteright{} amplifying
function can be represented by a diagonal matrix $\mathbf{W}\in\mathbb{C\mathit{^{L}}}$,
whose diagonal scalar entry $w_{l}^{\,}$ is the gain associated with
the $\mathit{l}$th selected forward relay antenna $\left(f\left(l\right),r\left(l\right)\right)$
(antenna $f\left(l\right)$ of relay $r\left(l\right)$). It should
be noted that previously defined $b\left(l\right)$ is not necessary
to be the same as $f\left(l\right)$ in the MIMO relay node. But they
have to belong to the same relay $r\left(l\right)$ because no cooperation
is operated among different relays. From (1), the amplified signal
is written as

\begin{equation}
\mathbf{t=Wy=WHx+Wn_{\textrm{r}}^{\,}}.\end{equation}

\noindent Usually, each relay node has an independent power supply,
the local power of which is bounded by $\mathit{P_{\textrm{loc}}}$
shown as

\begin{equation}
\varepsilon\left\{ tr\left(\mathbf{tt^{\textrm{H}}}\right)_{l,l}\right\} =\left(\mathbf{W^{\textrm{2}}}\left(\sigma_{x}^{2}\mathbf{HH^{\textrm{H}}+\mathbf{I_{\mathit{L}}}}\right)\right)_{l,l}\leq\mathit{P_{\textrm{loc}}}.\end{equation}

\noindent Assume full power in (4). Then $w_{l}^{\,}$ can be represented
by

\begin{equation}
w_{l}^{\,}=\sqrt{P_{\textrm{loc}}/\left(\sigma_{x}^{2}\left|\mathbf{h}_{\left(\mathit{b}(l),\mathit{r}(\mathit{l})\right)}\right|^{\textrm{2}}+1\right)}.\end{equation}

\noindent Under the condition of perfect synchronization in the relay
network, signal arriving at $\mathcal{D}$ is given by

\begin{equation}
\mathbf{z}=\mathbf{Gt+n_{\textrm{d}}^{\,}=GWHx+GWn_{\textrm{r}}^{\,}+n_{\textrm{d}}^{\,}},\end{equation}

\noindent where the noise term $\mathbf{n_{\textrm{d}}^{\,}\in\mathbb{C\mathit{^{N_{\textrm{d}}}}}}$
stands for the ZMCSCG vector at $\mathcal{D}$ with identity covariance
matrix. $\mathbf{G}\in\mathbb{C^{\mathit{N_{\textrm{d}}\times L}}}$
denotes the compound forward channel written as $\mathbf{G=\left[{\mathbf{g}}_{\left(\mathit{f}(\textrm{1}),\mathit{r}(\textrm{1})\right)},\cdots,{g}_{\left(\mathit{f}(\mathit{l}),\mathit{r}(\mathit{l})\right)},\cdots,{g}_{\left(\mathit{f}(\mathit{L}),\mathit{r}(\mathit{L})\right)}\right]}$.
By denoting equivalent channel ${\mathbf{H}}_{\textrm{eq}}^{\textrm{}}=\mathbf{GWH}$
and colored noise term ${\mathbf{n}}_{\textrm{eq}}^{\textrm{}}=\mathbf{GWn_{\textrm{r}}^{\,}+n_{\textrm{d}}^{\,}}$,
(6) can be further reduced to $\mathbf{\mathbf{z}=\mathbf{{\mathbf{H}}_{\textrm{eq}}^{\textrm{}}x}+{\mathbf{n}}_{\textrm{eq}}^{\textrm{}}}$.
According to \cite{[10]MMSE expression}, the MSE of symbol estimation
will achieve its minimum value when Wiener filter is employed. The
corresponding MSE is presented by

\begin{eqnarray}
Q & = & \sigma_{x}^{2}tr\left\{ \left({\mathbf{I}}\mathbf{_{\mathit{N_{\textrm{d}}}}}+\sigma_{x}^{\textrm{2}}{\mathbf{H}}_{\textrm{eq}}^{\textrm{}}{\mathbf{H}}_{\textrm{eq}}^{\textrm{H}}{\boldsymbol{\Phi}}^{-1}\right)^{-1}\right\} +\sigma_{x}^{\textrm{2}}\left(\mathit{N_{\textrm{s}}-N_{\textrm{d}}}\right)\nonumber \\
 & = & \sigma_{x}^{2}tr\left\{ \boldsymbol{\Phi}\left(\boldsymbol{\Phi}+\sigma_{x}^{\textrm{2}}{\mathbf{H}}_{\textrm{eq}}^{\textrm{}}{\mathbf{H}}_{\textrm{eq}}^{\textrm{H}}\right)^{-1}\right\} +\beta,\end{eqnarray}

\noindent where $\boldsymbol{\Phi}=\varepsilon\left\{ \mathbf{{\mathbf{n}}_{\textrm{eq}}^{\textrm{}}{\mathbf{n}}_{\textrm{eq}}^{\textrm{H}}}\right\} =\mathbf{GW\left(GW\right)^{\textrm{H}}}+{\mathbf{I}}\mathbf{_{\mathit{N_{\textrm{d}}}}}$
is the covariance matrix for ${\mathbf{n}}_{\textrm{eq}}^{\textrm{}}$.
Since $\beta=\sigma_{x}^{\textrm{2}}\left(\mathit{N_{\textrm{s}}-N_{\textrm{d}}}\right)$
is just a constant that is free from the minimization of $Q$, we
will omit $\beta$ hereafter.

\section{The Proposed Antenna Selection Algorithm}

Since (7) is a closed-form expression to evaluate the MSE of the system,
we can perform an exhaustive search to minimize $Q$ to obtain the
optimal antenna pair set. However, such a brute-force search is quite
infeasible considering the required trials could be as large as ${\displaystyle \sum}_{l=M}^{K}\textrm{C}_{K}^{l}\left({N}_{\textrm{r}}^{2}\right)^{l}$,
where ${N}_{\textrm{r}}^{2}$ refers to the number of candidate antenna
pairs at each relay node. In a modest case where $N_{\textrm{s}}=N_{\textrm{d}}=4$,
$N_{\textrm{r}}=2$, and $K=8$, the exhaustive search would involve
approximately $3.9\times10^{5}$ trials requiring the inversion of
a matrix of size $4\times4$ in each trial shown in (7). Moreover,
that figure would soon rocket to nearly 10 million if $K=10$. Therefore,
instead of approaching (7) directly, we investigate how the MSE is
affected when one more antenna pair is chosen so as to find a way
to optimize the system asymptotically.

Denote ${\mathbf{H}}_{l}=\mathbf{\left[{\mathbf{h}}_{\left(\mathit{b}(\textrm{1}),\mathit{r}(\textrm{1})\right)}^{\textrm{T}},\cdots,{\mathbf{h}}_{\left(\mathit{b}(\mathit{l}),\mathit{r}(\mathit{l})\right)}^{\textrm{T}}\right]^{\textrm{T}}}$
as the already selected backward channel, then the diagonal entries
of $\mathbf{W_{\mathit{l}}}$ can be obtained from (5). Next we turn
to the $\mathit{\left(l\textrm{+1}\right)}$th antenna pair. For every
un-selected candidate backward channel ${\mathbf{h}}_{m,k}$ of $\mathcal{S}$
to the $\mathit{m}$th antenna of the $\mathit{k}$th relay, we calculate
the associated relay gain $\mathit{w_{m,k}^{\textrm{}}}$ according
to (5) as

\begin{equation}
\mathit{w_{m,k}^{\textrm{}}}=\sqrt{P_{\textrm{loc}}/\left(\sigma_{x}^{\textrm{2}}\left|{\mathbf{h}}_{m,k}\right|^{\textrm{2}}+1\right)}.\end{equation}

\noindent Suppose that $\mathbf{G_{\mathit{l}}=\left[{\mathbf{g}}_{\left(\mathit{f}(\textrm{1}),\mathit{r}(\textrm{1})\right)},\cdots,{\mathbf{g}}_{\left(\mathit{f}(\mathit{l}),\mathit{r}(\mathit{l})\right)}\right]}$
is the selected partial forward channel. Let ${\mathbf{g}}_{n,k}$
be the $N_{\textrm{d}}\times1$ channel vector characterizing the
forward link from the $\mathit{n}$th antenna of the $\mathit{k}$th
relay to $\mathcal{\mathcal{D}}$. For simplicity, we use the notation
$\left(m,n,k\right)$ to represent the candidate antenna pair $\left(m,n\right)$
on the $k$th relay. Further denote

\noindent $\qquad\;{\boldsymbol{\Phi}}_{l}={\mathbf{I}}\mathbf{_{\mathit{N_{\textrm{d}}}}}+{\mathbf{G}}_{l}{\mathbf{W}}_{l}^{2}{\mathbf{G}}_{l}^{\textrm{H}},$

\noindent $\qquad\;{\mathbf{A}}_{\mathit{l}}={\boldsymbol{\Phi}}_{l}+\sigma_{x}^{\textrm{2}}\left({\mathbf{G}}_{l}{\mathbf{W}}_{l}{\mathbf{H}}_{l}\right)\left({\mathbf{G}}_{l}{\mathbf{W}}_{l}{\mathbf{H}}_{l}\right)^{\textrm{H}},$

\noindent $\qquad\;{\mathbf{F}}_{\mathit{m,n,k}}={\mathbf{G}}_{l}{\mathbf{W}}_{l}{\mathbf{H}}_{l}+\mathit{w_{m,k}^{\textrm{}}}{\mathbf{g}}_{n,k}^{\textrm{}}{\mathbf{h}}_{m,k}^{\textrm{}},$

\noindent $\qquad\;{\mathbf{u}}_{m,n,k}^{\,}=\sigma_{x}^{\textrm{2}}\left({\mathbf{G}}_{l}{\mathbf{W}}_{l}{\mathbf{H}}_{l}\right)\mathit{w_{m,k}^{\textrm{}}}{\mathbf{h}}_{m,k}^{\textrm{H}}+\mathit{w_{m,k}^{\textrm{2}}}{\mathbf{g}}_{n,k}^{\textrm{}},$

\noindent $\qquad\;{\boldsymbol{v}}_{\mathit{m,n,k}}^{\,}=\sigma_{x}^{\textrm{2}}{\mathbf{F}}_{\mathit{m,n,k}}\mathit{w_{m,k}^{\textrm{}}}{\mathbf{h}}_{m,k}^{\textrm{H}}.$

\noindent Our main result is summarized in the following proposition.
\begin{prop}
The MSE resulted from the $\mathit{\left(l\textrm{+1}\right)}$th
additional sub-channels ${\mathbf{h}}_{m,k}$, ${\mathbf{g}}_{n,k}$
with respect to the antenna pair $\left(m,n,k\right)$ is
\end{prop}
\begin{equation}
Q_{l+1}^{\left(m,n,k\right)}=\sigma_{x}^{2}tr\left\{ \left({\boldsymbol{\Phi}}_{l}+\mathit{w_{m,k}^{\textrm{2}}}{\mathbf{g}}_{n,k}^{\textrm{}}{\mathbf{g}}_{n,k}^{\textrm{H}}\right){\mathbf{C}}_{\left(m,n,k\right)}^{-1}\right\} ,\end{equation}

\noindent where ${\mathbf{C}}_{m,n,k}^{-1}={\mathbf{B}}_{m,n,k}^{-1}\mathit{-}\frac{{\mathbf{B}}_{m,n,k}^{-1}{\mathbf{g}}_{n,k}^{\textrm{}}{\boldsymbol{v}}_{\mathit{m,n,k}}^{\textrm{H}}{\mathbf{B}}_{m,n,k}^{-1}}{1+{\boldsymbol{v}}_{\mathit{m,n,k}}^{\textrm{H}}{\mathbf{B}}_{m,n,k}^{-1}{\mathbf{g}}_{n,k}^{\textrm{}}}$
and ${\mathbf{B}}_{m,n,k}^{-1}={\mathbf{A}}_{\mathit{l}}^{-1}\mathit{-}\frac{{\mathbf{A}}_{\mathit{l}}^{-1}{\mathbf{u}}_{m,n,k}^{\,}{\mathbf{g}}_{n,k}^{\textrm{H}}{\mathbf{A}}_{\mathit{l}}^{-1}}{1+{\mathbf{g}}_{n,k}^{\textrm{H}}{\mathbf{A}}_{\mathit{l}}^{-1}{\mathbf{u}}_{m,n,k}^{\,}}$.
\begin{proof}
According to (7) and by some mathematical manipulations, the MSE (with
$\beta$ omitted) with ${\mathbf{h}}_{m,k}$ and ${\mathbf{g}}_{n,k}$
added to $\mathbf{H_{\mathit{l}}}$ and $\mathbf{G_{\mathit{l}}}$
can be represented as

\smallskip{}

\noindent $Q_{l+1}^{\left(m,n,k\right)}$

\noindent $\;\;=\sigma_{x}^{2}tr\left\{ \begin{array}{c}
\left({\boldsymbol{\Phi}}_{l}+\mathit{w_{m,k}^{\textrm{2}}}{\mathbf{g}}_{n,k}^{\textrm{}}{\mathbf{g}}_{n,k}^{\textrm{H}}\right)\\
\times\left({\boldsymbol{\Phi}}_{l}+\mathit{w_{m,k}^{\textrm{2}}}{\mathbf{g}}_{n,k}^{\textrm{}}{\mathbf{g}}_{n,k}^{\textrm{H}}+\sigma_{x}^{\textrm{2}}{\mathbf{F}}_{\mathit{m,n,k}}{\mathbf{F}}_{\mathit{m,n,k}}^{\textrm{H}}\right)^{-1}\end{array}\right\} $

\noindent \begin{equation}
=\sigma_{x}^{2}tr\left\{ \begin{array}{c}
\left({\boldsymbol{\Phi}}_{l}+\mathit{w_{m,k}^{\textrm{2}}}{\mathbf{g}}_{n,k}^{\textrm{}}{\mathbf{g}}_{n,k}^{\textrm{H}}\right)\\
\times\left({\mathbf{A}}_{\mathit{l}}+{\mathbf{u}}_{m,n,k}^{\,}{\mathbf{g}}_{n,k}^{\textrm{H}}+{\mathbf{g}}_{n,k}^{\textrm{}}{\boldsymbol{v}}_{\mathit{m,n,k}}^{\textrm{H}}\right)^{-1}\end{array}\right\} .\end{equation}

\noindent By invoking the matrix inversion lemma \cite{[11]math book1},
that is \begin{equation}
\left(\mathbf{A+XY^{\textrm{H}}}\right)^{-1}=\mathbf{A}^{-1}\mathit{-}\mathbf{A}^{-1}\mathbf{X}\left(\mathbf{I}+\mathbf{Y}^{\textrm{H}}\mathbf{A}^{-1}\mathbf{X}\right)^{-1}\mathbf{Y}^{\textrm{H}}\mathbf{A}^{-1},\end{equation}

\noindent where $\mathbf{A}\in\mathbb{C^{\mathit{a\times a}}}$ and
$\mathbf{X},\mathbf{Y}\in\mathbb{C^{\mathit{a\times b}}}$. We can
evaluate (10) by means of a two-step recursion. Firstly, we denote
${\mathbf{B}}_{m,n,k}^{\,}={\mathbf{A}}_{\mathit{l}}+{\mathbf{u}}_{m,n,k}^{\,}{\mathbf{g}}_{n,k}^{\textrm{H}}$,
and calculate ${\mathbf{B}}_{m,n,k}^{-1}$ according to (11) as

\noindent \begin{equation}
{\mathbf{B}}_{m,n,k}^{-1}={\mathbf{A}}_{\mathit{l}}^{-1}\mathit{-}\frac{{\mathbf{A}}_{\mathit{l}}^{-1}{\mathbf{u}}_{m,n,k}^{\,}{\mathbf{g}}_{n,k}^{\textrm{H}}{\mathbf{A}}_{\mathit{l}}^{-1}}{1+{\mathbf{g}}_{n,k}^{\textrm{H}}{\mathbf{A}}_{\mathit{l}}^{-1}{\mathbf{u}}_{m,n,k}^{\,}}.\end{equation}

\noindent Then let ${\mathbf{C}}_{m,n,k}={\mathbf{B}}_{m,n,k}^{\,}+{\mathbf{g}}_{n,k}^{\textrm{}}{\boldsymbol{v}}_{\mathit{m,n,k}}^{\textrm{H}}$.
We have

\noindent \begin{equation}
{\mathbf{C}}_{m,n,k}^{-1}={\mathbf{B}}_{m,n,k}^{-1}\mathit{-}\frac{{\mathbf{B}}_{m,n,k}^{-1}{\mathbf{g}}_{n,k}^{\textrm{}}{\boldsymbol{v}}_{\mathit{m,n,k}}^{\textrm{H}}{\mathbf{B}}_{m,n,k}^{-1}}{1+{\boldsymbol{v}}_{\mathit{m,n,k}}^{\textrm{H}}{\mathbf{B}}_{m,n,k}^{-1}{\mathbf{g}}_{n,k}^{\textrm{}}}.\end{equation}

\noindent The proof is completed by substituting (13) into (10).
\end{proof}
As shown in (9), $Q_{l+1}^{\left(m,n,k\right)}$ can be evaluated
efficiently since ${\mathbf{A}}_{\mathit{l}}^{-1}$ is free from $\left(m,n,k\right)$
and only computed once for each $l$. Other computations involved
in (9) are no more than several vector/matrix multiplications. Based
on (9), we can iteratively activate more antenna pairs as long as
the corresponding MSE keeps decreasing. Although this approach will
not guarantee a global optimal solution as the exhaustive search,
it has the potential to find a good local optimal solution because
of three facts: (i) local optimality can be reflected in the non-increasing
MSE based searching; (ii) the noise terms have been correctly incorporated
into (9); (iii) $L$ can be as large as $K$ to exploit the diversity
gain of the network. Hence, we propose a MMSE based greedy antenna
selection algorithm summarized as follows.

\textit{}%
\begin{algorithm}
\textit{\caption{\textit{Greedy MSE Minimization (GMM)}}
}
\begin{enumerate}
\item Initialization: Set $l=0$, ${\mathbf{H}}_{l}$, ${\mathbf{G}}_{l}$,
${\mathbf{W}}_{l}=\slashed{O}$, ${\mathbf{A}}_{\mathit{l}}={\boldsymbol{\Phi}}_{\mathit{l}}={\mathbf{I}}\mathbf{_{\mathit{N_{\textrm{d}}}}}$,
${\mathit{previous\_MSE}}=+\infty$; Let $\Omega$ be the candidate
antenna pair set containing all $K{N}_{\textrm{r}}^{2}$ pairs $\left(m,n,k\right)$.
\item Iterative loop: Compute ${\mathbf{W}}_{l}$, ${\boldsymbol{\Phi}}_{l}$,
${\mathbf{A}}_{\mathit{l}}$ and ${\mathbf{A}}_{\mathit{l}}^{-1}$;
\\
For each antenna pair $\left(m,n,k\right)$ in $\Omega$, obtain
$\mathit{w_{m,k}^{\textrm{}}}$ according to (8). Then evaluate (9)
to get $Q_{l+1}^{\left(m,n,k\right)}$.
\item Select the $\left(l+1\right)$th antenna pair by: \\
$\left(b\left(j\right),f\left(j\right),r\left(j\right)\right)_{j=l+1}=\underset{\left(m,n,k\right)}{\arg\min}\left\{ \begin{array}{c}
Q_{l+1}^{\left(m,n,k\right)}\end{array}\right\} $.
\item If $\min\left\{ \begin{array}{c}
Q_{l+1}^{\left(m,n,k\right)}\end{array}\right\} <{\mathit{previous\_MSE}}$, \\
then ${\mathit{previous\_MSE}}$ $=\min\left\{ Q_{l+1}^{\left(m,n,k\right)}\right\} $;
\\
Set $l=l+1$; Eliminate the ${N}_{\textrm{r}}^{2}$ antenna pairs
associated with the relay $r\left(l\right)$ from $\Omega$ ; Update
${\mathbf{H}}_{l}=\left[{\mathbf{H}}_{l-1}^{\textrm{T}},{\mathbf{h}}_{\left(\mathit{b}(\mathit{l}),\mathit{r}(\mathit{l})\right)}^{\textrm{T}}\right]^{\textrm{T}}$,
${\mathbf{G}}_{l}=\left[{\mathbf{G}}_{l-1},{\mathbf{g}}_{\left(f(\mathit{l}),\mathit{r}(\mathit{l})\right)}^{\textrm{}}\right]$,
${\mathbf{W}}_{l}=\left[\begin{array}{cc}
{\mathbf{W}}_{l-1} & 0\\
0 & w_{b\left(l\right),r\left(l\right)}^{\textrm{}}\end{array}\right]$; Go to step 2.\\
Else, terminate with ${\mathbf{H}}_{l}$ and ${\mathbf{G}}_{l}$
as the selected backward channel and forward channel.
\end{enumerate}

\end{algorithm}

\section{Simulation Results And Discussions}

A distributed orthogonal relay selection (DORS) algorithm and a semi-orthogonization
(S-O) algorithm have been proposed in \cite{[8]DORS} and \cite{[9]S-O}
respectively. The DORS algorithm selects the antenna pair to maximize
the harmonic mean of dual-hop sub-channel gains achieved by ${\mathbf{h}}_{m,k}$
and ${\mathbf{g}}_{n,k}$, while in the S-O algorithm the authors
maximize the sum of the projection angles among the sub-channels of
${\mathbf{H}}_{l}$ and ${\mathbf{G}}_{l}$.

In our simulations, we adopt a realistic antenna setup as $N_{\textrm{s}}=N_{\textrm{d}}=M=4$,
$N_{\textrm{r}}=2$. The channels are assumed to be uncorrelated Rayleigh
fading, which are modeled as \textit{i.i.d.} ZMCSCG random variables
with unit covariance. Furthermore, we denote the receive SNR at the
relay nodes as ${SNR}_{1}=P_{\textrm{s}}$ (with the noise power be
normalized to 1). $P_{\textrm{loc}}$ is set to 5 dB above the noise
power and \textit{10000} Monte Carlo runs are conducted for each relay
deployment.

\begin{figure}[b]
\centering \includegraphics[width=8cm,height=2.5cm]{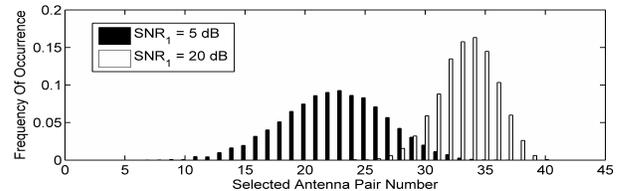} 
\vspace{-0.5em}
\renewcommand{\figurename}{Fig.}\caption{Frequency histogram of the selected antenna pairs (40 relays).}

\end{figure}

Firstly, we make a brief complexity comparison among the aforementioned
schemes. The exhaustive search scheme is nearly impossible to be analyzed
when $K>8$, which is verified by our simulation efforts. The DORS
and S-O algorithms stop the antenna selection procedure when $l=M$,
whereas the proposed GMM scheme tends to select more antennas ($M\leq l\leq K$)
until the MSE begins to increase. In each antenna selection loop,
although the GMM scheme is more involved than the DORS and S-O algorithms,
its implementation is feasible as explained earlier. Fig.\,2 presents
the frequency histogram of the selected antenna pairs for the GMM
scheme for $K=40$. It is not surprising to find that less antenna
pairs will be expected to participate the relaying when ${SNR}_{1}$
is smaller because the received signals at the relay nodes are more
likely to vanish under the noise floor. However, we do observe that
the GMM scheme will turn on more relay nodes than the DORS and S-O
algorithms, thus making the performance comparison unfair due to the
additional power gain. A simple way to separate the power gain from
the enhancement offered by the GMM scheme is to pose a global power
constraint on the selected $L$ forward antennas, \textit{i.e.}, instead
of allocating $P_{\textrm{loc}}$ for each activated relay node, we
dilute the relay power to $MP_{\textrm{loc}}/L$. Thereby, the total
power at the activated relay nodes will be the same for the DORS,
S-O and GMM schemes.

Fig.\,3 shows the MSE performance versus relay number for the DORS,
S-O, GMM and exhaustive search schemes. From Fig.\,3, we find that
the GMM scheme largely reduces the MSE compared to the DORS and S-O
algorithms, with the gain being more conspicuous when the noise issue
at the relay nodes becomes more serious (the upper set of curves,
${SNR}_{1}=5\,\mathcal{\textrm{dB}}$) and $K$ becomes larger. When
comparing the curves for the GMM algorithm with and without global
power constraint, we can draw the conclusion that the GMM algorithm
without the power bonus has already reaped most of the performance
gains. We also observe that the GMM scheme achieves the performance
close to that by exhaustive search, which further confirms the superiority
of the proposed scheme.

\begin{figure}[t]
\centering \includegraphics[width=8cm,height=6cm]{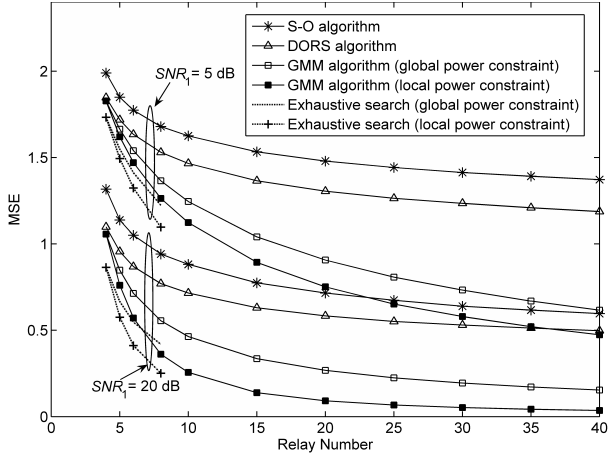}
\vspace{-0.5em}
\renewcommand{\figurename}{Fig.}\caption{MSE performance versus number of relay nodes $K$.}

\end{figure}

To illustrate how the MSE performance gains in Fig.\,3 are interpreted
into BER decrease, we plot the average BER curves in Fig.\,4 with
${SNR}_{1}$ varying from 0 dB to 30 dB. For all SNR cases, we deploy
15 relays. In addition, we assume that the Wiener filter is employed
as the symbol detection filter, and the symbols are obtained from
the QPSK constellation. As seen from Fig.\,4, the proposed GMM scheme
shows much steeper BER slope, indicating that more diversity is exploited
in the system. With respect to the error floor caused by the limited
power in the second hop channels, the proposed GMM scheme achieves
considerable gains in orders.

\begin{figure}[t]
\centering \includegraphics[width=8cm,height=6cm]{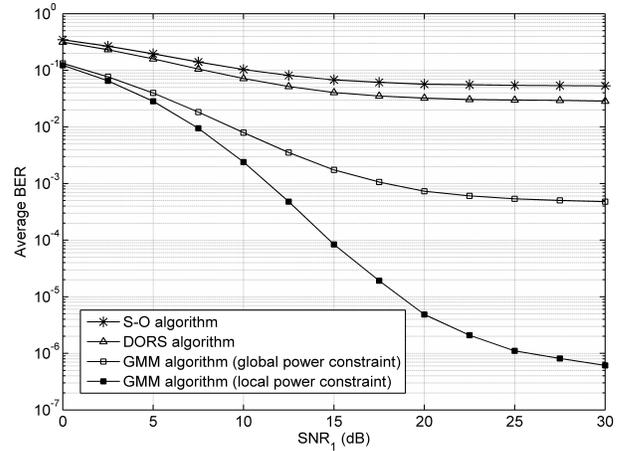}
\vspace{-0.5em}
\renewcommand{\figurename}{Fig.}\caption{BER performance versus \textit{${SNR}_{1}$} (dB).}

\end{figure}

\section{Conclusion}

In this paper, we proposed a greedy antenna selection algorithm to
minimize the MSE resulted from selecting an additional relay antenna
of an AF MIMO relay system. To reduce the complexity, the antenna
selection process is carried out iteratively. Simulation results show
that the proposed scheme exhibits much better performance than the
existing schemes in terms of MSE and BER, and the gain is more pronounced
in noisy channels.

\end{document}